\magnification\magstep1

\font\man=manfnt
\setbox0=\hbox{\man R}
\newdimen\Blankpix \Blankpix=\wd0

\input picmac
\catcode`\@=11
\def\squine(#1,#2,#3,#4,#5,#6){\setbox\@picbox\hbox{\fiverm.}%
 \global\@xoldpt=#1\unitlength \global\@yoldpt=#4\unitlength \kern\@xoldpt
 \@xi=\@xoldpt \@xii=#2\unitlength \@xiii=#3\unitlength
 \@yi=\@yoldpt \@yii=#5\unitlength \@yiii=#6\unitlength
 \squinerec
 \@xpt=#3\unitlength \@ypt=#6\unitlength \@addpoint
 \raise\@ypt\copy\@picbox}
\def\testnear#1#2{\@save=#1\advance\@save-#2%
 \ifdim\@save<\z@ \@save=-\@save\fi \ifdim\@save>.2\p@ \fartrue \fi}
\catcode`\@=12

\newcount\xxx \newcount\yyy
\def\uu{\put(\the\xxx,\the\yyy){\line(0,1)1}\advance\yyy1}
\def\dd{\put(\the\xxx,\the\yyy){\line(0,-1)1}\advance\yyy-1}
\def\ll{\put(\the\xxx,\the\yyy){\line(-1,0)1}\advance\xxx-1}
\def\rr{\put(\the\xxx,\the\yyy){\line(1,0)1}\advance\xxx1}

\font\ninerm=cmr9
\font\logo=logo10 
\font\logosl=logosl10 
\def\MF{{\logo META}\-{\logo FONT}}
\def\MFbook{{\sl The {\logosl METAFONT}\kern1pt book}}

\def\bib{\par\noindent\hangindent 20pt}
\def\pfbox
  {\hbox{\hskip 3pt\lower1pt\vbox{\hrule
  \hbox to 5pt{\vrule height 6pt\hfill\vrule}
  \hrule}}\hskip3pt}

\centerline{\bf A note on digitized angles}\footnote{}{The preparation
of this note was supported in part by National Science Foundation
grant grant CCR--8610181.}
\centerline{Donald E. Knuth, Stanford University}
\bigskip
{\narrower\smallskip\noindent
{\it Abstract}. 
We study the configurations of pixels that occur when two digitized
straight lines meet each other.
\smallskip}

\bigskip
About ten years ago I was supervising the Ph.D. thesis of 
Chris Van Wyk~[4],
which introduced the {\ninerm IDEAL} language for describing 
pictures~[5].
Two of his example illustrations showed arrows constructed from
straight lines something like this:

$$\unitlength=1mm
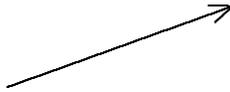
\beginpicture(25,9)(0,0)
\put(0,0){\squine(0,12.5,25,0,4.5,9)}
\put(0,0){\squine(22.3459,23.6730,25,9.1253,9.0626,9)}
\put(0,0){\squine(23.0347,24.0174,25,7.2118,8.1059,9)}
\endpicture
$$

\noindent
When I looked at them, I was sure that there must be a bug either in
{\ninerm IDEAL} or in the {\ninerm TROFF} processor that typeset the
{\ninerm IDEAL} output, because the long shafts of the arrows did not
properly bisect the angle made by the two short lines of the
arrowheads.
The shafts seemed to be drawn one pixel too high or too low. Chris
spent many hours together with Brian Kernighan trying to find out what
was wrong, but no errors could be pinned down. Eventually his thesis was
printed on a high-resolution phototypesetter, and the problem became
much less noticeable than it had been on the laser-printed proofs.
There still was a glitch, but I~decided not to hold up Chris's
graduation for the sake of a misplaced pixel.

I remembered this incident at the end of 1983, when I was getting
ready to write a new version of the \MF\ system for digital 
art~[3].
I~didn't want my system to have such a flaw. But to my surprise,
I~learned that the problem is actually unavoidable in raster output:
It is almost impossible to bisect a digitized angle exactly, except in
very special circumstances. The two ``halves'' of the angle will
necessarily appear somewhat different from each other, unless the
resolution is quite high. Therefore Van Wyk (and Kernighan) were
vindicated.  Similar problems are bound to occur in MacDraw and in any
other drawing package.

For example, one of the things I noticed was the
following curious fact. Consider the $45^{\circ}$ angle that is made
when a straight line segment 
of slope~2 comes up to a point $(x_0,y_0)$ and then
goes down along another line of slope~$-3$:

$$\unitlength=4pt
\beginpicture(19,13)(-10,-8)
\put(0,0){\line(1,-3){3}}
\put(0,0){\line(-1,-2){4}}
\put(-5,.4){\makebox(0,0){$(x_0,y_0)$}}
\put(-13,-6.5){\makebox(0,0){$(x_0+2t,\,y_0-t)$}}
\put(12.7,-7.6){\makebox(0,0){$(x_0+u,\,y_0-3u)$}}
\endpicture
$$

\noindent
If we digitize this angular path, the result will take one of five
different shapes, depending on the value of 
the intersection point $(x_0,y_0)$, whose coordinates are not
necessarily integers. The possibilities    are
$$\unitlength=4pt
P_0:\ \vcenter
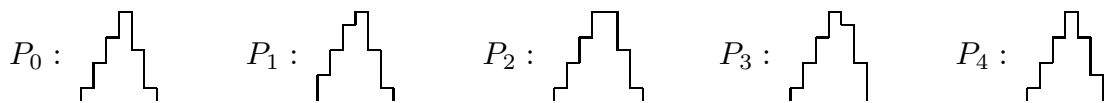
{\hbox{\beginpicture(8,7)(0,0)\xxx=0 \yyy=0
 \uu\rr\uu\uu\rr\uu\uu\rr\uu\uu\rr\dd\dd\dd\rr\dd\dd\dd\rr\dd\endpicture}}
\qquad
P_1:\ \vcenter{\hbox{\beginpicture(8,7)(0,0)\xxx=0 \yyy=0
 \uu\uu\rr\uu\uu\rr\uu\uu\rr\uu\rr\dd\dd\dd\rr\dd\dd\dd\rr\dd\endpicture}}
\qquad
P_2:\ \vcenter{\hbox{\beginpicture(8,7)(0,0)\xxx=0 \yyy=0
 \uu\rr\uu\uu\rr\uu\uu\rr\uu\uu\rr\rr\dd\dd\dd\rr\dd\dd\dd\rr\dd\endpicture}}
\qquad
P_3:\ \vcenter{\hbox{\beginpicture(8,7)(0,0)\xxx=0 \yyy=0
 \uu\rr\uu\uu\rr\uu\uu\rr\uu\uu\rr\dd\rr\dd\dd\dd\rr\dd\dd\dd\endpicture}}
\qquad
P_4:\ \vcenter{\hbox{\beginpicture(8,7)(0,0)\xxx=0 \yyy=0
 \uu\rr\uu\uu\rr\uu\uu\rr\uu\uu\rr\dd\dd\rr\dd\dd\dd\rr\dd\dd\endpicture}}
$$

\smallskip
\noindent
If this $45^{\circ}$ angle provides the left half of an arrowhead, the
right half of the arrowhead will be a~$45^{\circ}$ angle made by a line of
slope~$-3$ meeting a line of slope~$-1/2$,

\vskip-10pt

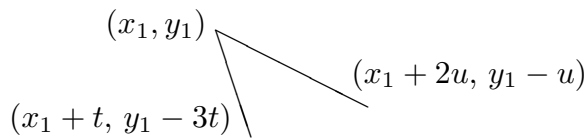
$$\unitlength=4pt
\beginpicture(19,13)(-6,-8)
\put(0,0){\line(1,-3){2.8}}
\put(0,0){\line(2,-1){12}}
\put(-4.7,.4){\makebox(0,0){$(x_1,y_1)$}}
\put(20,-3.5){\makebox(0,0){$(x_1+2u,\,y_1-u)$}}
\put(-7.5,-7){\makebox(0,0){$(x_1+t,\,y_1-3t)$}}
\endpicture
$$

\noindent
For this angle there are, similarly,
 five possibilities after digitization, namely

$$\unitlength=4pt
Q_0:\ \vcenter
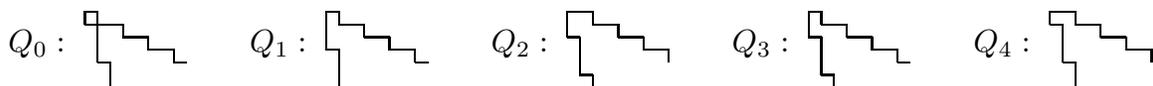
{\hbox{\beginpicture(8,7)(0,0)\xxx=2 \yyy=0
 \uu\uu\ll\uu\uu\uu\ll\uu\rr\dd\rr\rr\dd\rr\rr\dd\rr\rr\dd\rr\endpicture}}
\qquad
Q_1:\ \vcenter{\hbox{\beginpicture(8,7)(0,0)\xxx=1 \yyy=0
 \uu\uu\uu\ll\uu\uu\uu\rr\dd\rr\rr\dd\rr\rr\dd\rr\rr\dd\rr\endpicture}}
\qquad
Q_2:\ \vcenter{\hbox{\beginpicture(8,7)(0,0)\xxx=2 \yyy=0
 \uu\ll\uu\uu\uu\ll\uu\uu\rr\rr\dd\rr\rr\dd\rr\rr\dd\rr\rr\dd\endpicture}}
\qquad
Q_3:\ \vcenter{\hbox{\beginpicture(8,7)(0,0)\xxx=2 \yyy=0
 \uu\ll\uu\uu\uu\ll\uu\uu\rr\dd\rr\rr\dd\rr\rr\dd\rr\rr\dd\rr\endpicture}}
\qquad
Q_4:\ \vcenter{\hbox{\beginpicture(8,7)(0,0)\xxx=2 \yyy=0
 \uu\uu\ll\uu\uu\uu\ll\uu\rr\rr\dd\rr\rr\dd\rr\rr\dd\rr\rr\dd\endpicture}}
$$

\smallskip

To complete the arrowhead, we should match the left angle $P_i$ with
an appropriate~$Q_j$. But
none of the~$Q$'s has the same shape as
 any of the~$P$'s. And this is the point:
Human eyes tend to judge the magnitude of an angle by its appearance
at the tip. By this criterion, some of these angles appear to be quite
a bit larger than others (except at high resolutions). Hence it
is not surprising that a correctly drawn angle of type~$P$ would
appear to be unequal to a correctly drawn angle of type~$Q$, even
though both angles would really be~$45^{\circ}$ when drawn with
infinite resolution. (The patterns of white pixels, not black pixels,
are the source of the inconsistency.) Here, for example, are four
quite properly digitized arrows with shafts of increasing thickness:

$$\catcode`Q=\active \defQ{\kern\Blankpix}
\catcode`5=\active \def5{Q}
\catcode`6=\active \def6{Q}
\catcode`7=\active \def7{Q}
\def\arrow{
\vcenter{\man\offinterlineskip\halign{\hbox{##}\hfil\cr
QQQQQQR\cr
QQQQQRRRR\cr
QQQQQRRRRRR\cr
QQQQRRRRRRRRR\cr
QQQQRRRRR7RRRRR\cr
QQQRRR5RR6QQRRRRR\cr
QQQRRQ6RR5QQQQRRRRR\cr
QQRRRQ7RRR7QQQQQRR\cr
QQRRQQQ5RR6\cr
QRRRQQQ6RR5\cr
QRRQQQQ7RRR7\cr
RRRQQQQQ5RR6\cr
RRQQQQQQ6RR5\cr
QQQQQQQQ7RRR7\cr
QQQQQQQQQ5RR6\cr
QQQQQQQQQ6R\cr}}}
\arrow\qquad
\def5{R}\arrow\qquad
\def6{R}\arrow\qquad
\def7{R}\arrow
$$

We might also want to know the probability that the digitized shape
will be of a particular type~$P_k$, when the corner point $(x_0,y_0)$
is chosen at random in the plane. Is one of the patterns more likely
to occur than the others? The answer is no, when we use the most
natural method of digitization; each $P_k$ will be
obtained with probability 1/5. Similarly, each of the five
shapes~$Q_k$ turns out to be equally likely, as $(x_1,y_1)$ varies.

The main purpose of this note is to prove that the facts just stated
are special cases of a general phenomenon:

\proclaim Theorem. When a line of slope $a/b$ meets a line of slope
$c/d$ at a point $(x_0,y_0)$, the number of different digital shapes
it can produce as $(x_0,y_0)$ varies is $\vert ad-bc\vert$. Moreover,
each of these shapes is equally likely to occur, if $(x_0,y_0)$ is
chosen uniformly in the plane.

\noindent
We assume that $a/b$ and $c/d$ are rational numbers in lowest terms.
Two digitized shapes are considered to be equal if they are identical
after translation; rotation and reflection are not allowed.

Before we can prove the theorem, we need to define exactly what it
means to digitize a curve. For this, we follow the general idea
explained, for example, in~[3, Chapter~24]. We consider the plane to
be tiled with pixels, which are the unit squares whose corners have
integer coordinates. Our goal is to modify a given curve so that it
travels entirely on the boundaries between pixels. If
 the curve is given in parametric form by the function
$z(t)=\bigl(x(t),y(t)\bigr)$ as $t$~varies, its digitization is
essentially defined by the formula
$${\rm round}\;z(t)=\bigr({\rm round}\;x(t),\,{\rm round}\;
y(t)\bigr)$$
as $t$ varies, where  round$(\alpha)$  is the integer
nearest~$\alpha$. 

We need to be careful, of course, when rounding values that are
halfway between integers, because round$(\alpha)$ is undefined
in such cases. Let us assume for convenience that the path $z(t)$ does
not go through any pixel centers; i.e., that $z(t)$ is never equal to
$(m+{1\over 2},n+{1\over 2})$ for integer~$m$ and~$n$. (Exact hits on
pixel centers occur with probability zero, so they can be ignored in
the theorem we wish to prove. An infinitesimal shift of the path can
be used to avoid pixel centers in general, 
therefore avoiding the ambiguities pointed out in Bresenham's
interesting discussion~[1]; but we need not deal with
such complications.) Under this assumption, whenever we have
$x(t)=m+{1\over 2}$ so that `round~$x(t)$'  is ambiguous, the
value of  round~$y(t)=n$ will be unambiguous, and we can
include the entire line segment from $(m,n)$ to $(m+1,n)$ in the
digitized path. Similarly, when $t$ reaches a value such that
round~$x(t)=m$ but  round~$y(t)=n$ 
or $n+1$, we include the
entire segment from $(m,n)$ to $(m,n+1)$. This convention defines the
desired digitized path, round~$z(t)$.

When the path $z(t)$ returns to its starting point or begins and ends
at infinity, without intersecting itself, it defines a region in
the plane. The corresponding digitized path,  round~$z(t)$, also
defines a region; and this digitized region turns out to have a simple
characterization, when we apply standard mathematical 
conventions about ``winding
numbers'': {\sl The pixel with corners at\/ $(m,n)$, $(m+1,n)$,
$(m,n+1)$, $(m+1,n+1)$ belongs to the digitized region defined by\/
{\rm round}~$z(t)$ if and only if its center point\/ 
$(m+{1\over 2},
\allowbreak
n+{1\over 2})$ belongs to the undigitized region defined by\/
$z(t)$}.  (This beautiful property of digital curves is fairly easy to
verify in simple cases, but a rigorous proof is  difficult
because it relies ultimately on things like the Jordan Curve Theorem.
The necessary details appear in an appendix to John Hobby's
thesis~[2], Theorem A.4.1.)

Now we are ready to begin proving the desired result. The region
defined by an angle at $(x_0,y_0)$ with lines of slopes $a/b$ and
$c/d$ can be characterized by the inequalities
$$a(x-x_0)-b(y-y_0)\geq 0\,;\qquad c(x-x_0)-d(y-y_0)\geq 0\,.$$
$\bigl($We may need to reverse the signs, depending on which of the four
regions defined by two lines through $(x_0,y_0)$ are assumed to be defined
by the given angular path; this can be done by changing $(a,b)$ to
$(-a,-b)$ and/or $(c,d)$ to $(-c,-d)$.$\bigr)$
This region contains the pixel with lower left corner $(m,n)$ if and
only if 
$$\textstyle a(m+{1\over 2}-x_0)-b(n+{1\over 2}-y_0)\geq 0\,;\qquad
c(m+{1\over 2}-x_0)-d(n+{1\over 2}-y_0)\geq 0\,.$$
We can simplify the notation by combining several constants,
letting
$\alpha =a(x_0-{1\over
2})-b(y_0-{1\over 2})$ and $\beta=c(x_0-{1\over 2})-d(y_0-{1\over 2})$:
$$am-bn\geq\alpha\,;\qquad cm-dn\geq\beta\,.$$
Let $R(\alpha,\beta)$ be the digitized region consisting of all
integer pairs $(m,n)$ satisfying this condition; these are the pixels
in the digitized angle corresponding to $(x_0,y_0)$.

As noted above, it is safe to assume that the pixel centers do not
exactly touch the lines forming the angle; thus we are free to
stipulate that $am-bn\not=\alpha$ and $cm-dn\not=\beta$ for all pairs
of integers $(m,n)$. However, if equality does occur, we might as well
define the digitized region $R(\alpha,\beta)$ by the general
inequalities $am-bn\geq \alpha$ and $cm-dn\geq \beta$, 
as stated, instead of treating this circumstance as a special case.
Notice that $R(\alpha,\beta)$ is equal to $R(\lceil\beta\rceil)$;
therefore we can assume that~$\alpha$ and~$\beta$ are integers in the
following discussion. 

Another corner point $(x'_0,y'_0)$ will lead to parameters
$(\alpha',\beta')$ defining another region $R(\alpha',\beta')$ in the
same way. The two regions $R(\alpha,\beta)$ and $R(\alpha',\beta')$
have the same shape if and only if one is a translation of the other; i.e.,
$R(\alpha,\beta)\equiv R(\alpha',\beta')$ if and only if there exist
integers $(k,l)$ such that
$$(m,n)\in R(\alpha,\beta)\Longleftrightarrow (m-k,n-l)\in
R(\alpha',\beta')\,.$$
Our main goal is to prove that the number of distinct region shapes,
according to this notion of equivalence, is exactly $\vert ad-bc\vert$.

\proclaim Lemma. Let $\alpha$, $\beta$, $\alpha'$, $\beta'$ be
integers. Then $R(\alpha,\beta)\equiv R(\alpha',\beta')$ with respect
to slopes $a/b$ and $c/d$ if and only if $\alpha-\alpha'=ka-lb$ and
$\beta-\beta'=kc-ld$ for some integers $(k,l)$.

\noindent
{\it Proof}.
Assume that $R(\alpha,\beta)\equiv R(\alpha',\beta')$ with respect to
$a/b$ and $c/d$, and let $(k,l)$ be the corresponding translation
amounts. Thus we have
$$\left\{{am-bn\geq\alpha\atop cm-dn\geq\beta}\right\}\;\Longleftrightarrow \;
\left\{{a(m-k)-b(n-l)\geq\alpha'\atop 
c(m-k)-d(n-l)\geq\beta'}\right\}$$
for all integer pairs $(m,n)$. 
Let $\alpha''=\alpha'+ka-lb$ and $\beta''=\beta'+kc-ld$, so that
$$\left\{{am-bn\geq\alpha\atop c,-dn\geq\beta}\right\}
\;\Longleftrightarrow \;
\left\{{am-bn\geq\alpha''\atop
cm-dn\geq\beta''}\right\}$$
for all integers $(m,n)$. This implies that $\alpha=\alpha''$ and
$\beta=\beta''$. For if, say, we had $\alpha<\alpha''$, we could find
integers~$m$ and~$n$ such that $am-bn=\alpha$ and $cm-dn\geq\beta$,
because $a$ and~$b$ are relatively prime; this would satisfy the
inequalities on the left but not on the right. (More precisely, we
could use Euclid's algorithm to find integers~$a'$ and~$b'$ such that
$aa'-bb'=1$. Then the values $(m,n)=(\alpha a'+bx,\,\alpha b'+ax)$
would satisfy the left inequalities but not the right, for infinitely
many integers~$x$, because $ad-bc\neq 0$.)

Thus $R(\alpha,\beta)\equiv R(\alpha',\beta')$ implies that
$\alpha-\alpha'=ka-lb$ and $\beta-\beta'=kc-ld$. The converse is
trivial.\quad\pfbox

\medskip
Let $k$ and $l$ be integers such that $\alpha=ka-lb$. The lemma tells
us that $R(\alpha,\beta)\equiv R(0,\beta-kc+ld)$; hence every
digitized region $R(\alpha,\beta)$ has the same shape as some
digitized region $R(\alpha',\beta')$ in which $\alpha'=0$.

It remains to count the inequivalent regions $R(0,\beta)$ when $\beta$
is an integer. According to the lemma we have $R(0,\beta)\equiv
R(0,\beta')$ if and only if there exist integers $(k,l)$ with
$0=ka-lb$ and $\beta-\beta'=kc-ld$. But $ka=lb$ if and only if $k=bx$
and $l=ax$ for some integer~$x$; hence the condition reduces to
$\beta-\beta'=bxc-axd=x(bc-ad)$. In other words, $R(0,\beta)\equiv
R(0,\beta')$ if and only if $\beta-\beta'$ is a multiple of $ad-bc$.
The number of inequivalent regions is therefore $\vert ad-bc\vert$, as
claimed. 

To complete the proof of the theorem, we must also verify that each of
the equivalence classes is equally likely to be the class of the
digitized angular region, when the intersection point $(x_0,y_0)$ is
chosen at random in the plane. The notational change from $(x_0,y_0)$
to $(\alpha,\beta)$ maps equal areas into equal areas; so we want to
prove that the equivalence class of $R(\alpha,\beta)$ is uniformly
distributed among the $\vert ad-bc\vert$ possibilities, when the real
numbers $(\alpha,\beta)$ are chosen at random. Choosing real numbers
$(\alpha,\beta)$ at random leads to uniformly distributed pairs of
integers $(\lceil\alpha\rceil,\lceil\beta\rceil)$. 
And if $\lceil\alpha\rceil$ has
any fixed value and $\lceil\beta\rceil$ runs through all integers, the
equivalence class of $R(\alpha,\beta)$ runs cyclically through all
$\vert ad-bc\vert$ possibilities.

Q.E.D.\quad\pfbox

\medskip
A close inspection of this proof shows that we can give explicit
formulas for the sets of intersection points $(x_0,y_0)$ that produce
equivalent shapes. Let $D=\vert ad-bc\vert$ and let $R_j$~denote the
shape corresponding to region $R(0,j)$ in the proof, where $0\leq
j<D$. Then the digitized angle will have shape~$R_j$ if and only if
$(x_0,y_0)$ lies in the parallelogram whose corners are $\bigl({1\over
2}-bj/D,{1\over 2}-aj/D\bigr)$ plus
$$\bigl((b-d)/D,(a-c)/D\bigr)\,,\quad (-d/D,-c/D)\,,\quad
(b/D,a/D)\,,\quad (0,0)\,,$$
or in a parallelogram obtained by shifting this one by an integer
amount $(m,n)$. 

In the special case $a/b=2/1$ and $c/d=-3/1$, the shapes~$R_j$ are
what we called~$P_j$ above; in the special case $a/b=3/(-1)$ and
$c/d=-1/2$, the $R_j$ are what we called~$Q_j$. The shapes that appear
in the digitized angles depend on the values of $(x_0\bmod 1, y_0\bmod
1)$ in the unit square, according to the following diagrams:
$$\unitlength=.5mm
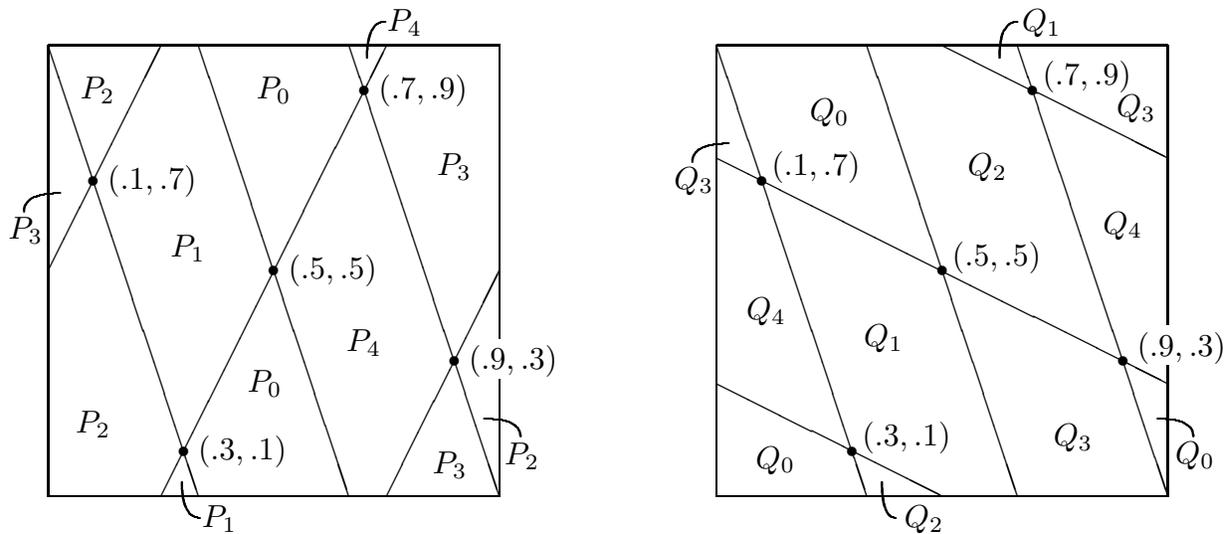
\beginpicture(120,120)(-10,-5)
\put(0,0){\line(0,1){100}}
\put(0,100){\line(1,0){100}}
\put(100,0){\line(0,1){26}}
\put(100,35){\line(0,1){65}}
\put(0,0){\line(1,0){100}}
\put(10,70){\disk{2}} \put(23,70){\makebox(0,0){$(.1,.7)$}}
\put(30,10){\disk{2}} \put(43,10){\makebox(0,0){$(.3,.1)$}}
\put(50,50){\disk{2}} \put(63,50){\makebox(0,0){$(.5,.5)$}}
\put(70,90){\disk{2}} \put(83,90){\makebox(0,0){$(.7,.9)$}}
\put(90,30){\disk{2}} \put(103,30){\makebox(0,0){$(.9,.3)$}}
\put(33.3333,0){\line(-1,3){33.3333}}
\put(66.6667,0){\line(-1,3){33.3333}}
\put(100,0){\line(-1,3){33.3333}}
\put(0,50){\line(1,2){25}}
\put(25,0){\line(1,2){50}}
\put(75,0){\line(1,2){25}}
\put(-5,59){\makebox(0,0){$P_3$}}
\put(11,90){\makebox(0,0){$P_2$}}
\put(31,55){\makebox(0,0){$P_1$}}
\put(38,-5){\makebox(0,0){$P_1$}}
\put(10,16){\makebox(0,0){$P_2$}}
\put(50,90){\makebox(0,0){$P_0$}}
\put(48,25){\makebox(0,0){$P_0$}}
\put(79,105){\makebox(0,0){$P_4$}}
\put(70,34){\makebox(0,0){$P_4$}}
\put(90,73){\makebox(0,0){$P_3$}}
\put(89,7){\makebox(0,0){$P_3$}}
\put(105,9){\makebox(0,0){$P_2$}}
\put(0,0){\squine(29,29,33,3,-5,-5)}
\put(0,0){\squine(70,70,74,97,105,105)}
\put(0,0){\squine(-6,-6,2,63,67,67)}
\put(0,0){\squine(104,104,96,13,17,17)}
\endpicture
\qquad\qquad
\beginpicture(120,120)(-10,-5)
\put(0,0){\line(0,1){100}}
\put(0,100){\line(1,0){100}}
\put(100,0){\line(0,1){29}}
\put(100,38){\line(0,1){62}}
\put(0,0){\line(1,0){100}}
\put(10,70){\disk{2}} \put(22,73){\makebox(0,0){$(.1,.7)$}}
\put(30,10){\disk{2}} \put(42,13){\makebox(0,0){$(.3,.1)$}}
\put(50,50){\disk{2}} \put(62,53){\makebox(0,0){$(.5,.5)$}}
\put(70,90){\disk{2}} \put(82,93){\makebox(0,0){$(.7,.9)$}}
\put(90,30){\disk{2}} \put(103,33){\makebox(0,0){$(.9,.3)$}}
\put(33.3333,0){\line(-1,3){33.3333}}
\put(66.6667,0){\line(-1,3){33.3333}}
\put(100,0){\line(-1,3){33.3333}}
\put(50,0){\line(-2,1){50}}
\put(100,25){\line(-2,1){100}}
\put(100,75){\line(-2,1){50}}
\put(-5,70){\makebox(0,0){$Q_3$}}
\put(11,41){\makebox(0,0){$Q_4$}}
\put(13,8){\makebox(0,0){$Q_0$}}
\put(25,85){\makebox(0,0){$Q_0$}}
\put(46,-5){\makebox(0,0){$Q_2$}}
\put(37,35){\makebox(0,0){$Q_1$}}
\put(79,13){\makebox(0,0){$Q_3$}}
\put(60,73){\makebox(0,0){$Q_2$}}
\put(72,105){\makebox(0,0){$Q_1$}}
\put(93,86){\makebox(0,0){$Q_3$}}
\put(90,60){\makebox(0,0){$Q_4$}}
\put(106,9){\makebox(0,0){$Q_0$}}
\put(0,0){\squine(36,36,39,3,-5,-5)}
\put(0,0){\squine(62,62,66,97,105,105)}
\put(0,0){\squine(-6,-6,2,74,78,78)}
\put(0,0){\squine(105,105,97,13,17,17)}
\endpicture
$$

\noindent
Notice that the parallelograms ``wrap around'' modulo 1, each taking
up an area of~1/5.

When the slope of either line forming an angle is irrational, the
number of possible shapes is infinite (indeed, uncountable). But we can
still study such digitizations by investigating the shape only in the
immediate neighborhood of the intersection point; after all, those
pixels are the most critical for human perception. For example,
exercises 24.7--9 of 
\MFbook~[3]
discuss the proper way to adjust the vertices of an equilateral
triangle so that it will digitize well.

The moral of this story, assuming that stories ought to have a moral,
is probably this: If you want to bisect an angle in such a way that
both halves of the bisected angle are visually equivalent, then the
line of bisection should be such that reflections about this line
always map
pixels into pixels. Thus, the bisecting line should be horizontal or
vertical or at a $45^{\circ}$~diagonal, and it should pass through
pixel corners and/or pixel centers. Furthermore, your line-rendering
algorithm should produce symmetrical results about the line of
reflection (see~[1]).

This subject is clearly ripe for a good deal of further investigation.

\bigskip\noindent
{\bf Acknowledgments.}\enspace I wish to thank the referees and the
editor for their comments. In particular, one of the referees
suggested the present proof of the theorem; my original version was
much more complicated.

\bigskip
\centerline{\bf Bibliography}
\medskip

\bib
[1]\enspace
Jack E. Bresenham, ``Ambiguities in incremental line rastering,''
{\sl IEEE Computer Graphics and Applications\/ \bf 7}, 5 (May 1987),
31--43.

\bib
[2]\enspace
John Douglas Hobby, Digitized brush trajectories. Ph.D. thesis,
Stanford University, April 1985. Also published as report
STAN-CS-85-1070. 

\bib
[3]\enspace
Donald E. Knuth, \MFbook, Volume~C of {\sl Computers \&
Typesetting}, (Reading, Mass.: Addison\kern.1em--Wesley, 1986).

\bib
[4]\enspace 
Christopher John Van Wyk, A language for typesetting graphics.
Ph.D. thesis, Stanford University, June 1980. Also published as report
STAN-CS-80-803. The ``arrow'' illustrations that prompted this
research appear on page~20.

\bib
[5]\enspace
Christopher J. Van Wyk, ``A graphics typesetting language,'' {\sl
SIGPLAN Notices\/ \bf 16}, 6 (June 1981), 99--107.

\bye